# A combined model for pseudorapidity distributions in Cu-Cu collisions at BNL-RHIC energies

Z. J. Jiang[*], J. Wang, Y. Huang

*College of Science, University of Shanghai for Science and Technology, Shanghai 200093, China*

The charged particles produced in nucleus-nucleus collisions come from leading particles and those frozen out from the hot and dense matter created in collisions. The leading particles are conventionally supposed having Gaussian rapidity distributions normalized to the number of participants. The hot and dense matter is assumed to expand according to the unified hydrodynamics, a hydro model which unifies the features of Landau and Hwa-Bjorken model, and freeze out into charged particles from a space-like hypersurface with a proper time of $\tau_{\text{FO}}$. The rapidity distribution of this part of charged particles can be derived out analytically. The combined contribution from both leading particles and unified hydrodynamics is then compared against the experimental data performed by BNL-RHIC-PHOBOS Collaboration in different centrality Cu-Cu collisions at $\sqrt{s_{\text{NN}}} = 200$ and 62.4 GeV, respectively. The model predictions are in well consistent with experimental measurements.

*Keywords*: Unified hydrodynamics; pseudorapidity distribution; leading particles.

PACS number(s): 25.75.Ag, 25.75.Ld, 24.10.Nz

1**. Introduction**

The application of relativistic hydrodynamics to high energy physics may be traced back to the pioneering work of Landau in 1953 [1]. In recent years, the most important achievement on this topic is the discovery that the spatiotemporal evolution of matter created in high energy heavy ion or particle collisions possesses the features of collective flow with strong interaction and behaves nearly like an ideal fluid with a very little viscosity [2-25].

Owing to the high degree of nonlinearity and interconnection of hydro equations, it has been being a formidable task to solve them analytically. This is the reason why from the times of Landau till now this problem is only limited to 1+1 expansion for ideal fluid with simple equation of state. The expansions for higher dimensions or situations including viscosity have little analytical discussion. The general exact solution for cases such as these is far from being obtained so far. The treatment of these problems usually resorts to Monte Carlo simulations. In Monte Carlo simulations, beside a powerful calculation system, there also needs a sophisticated skill for avoiding instabilities in solving partial differential hydro equations. Furthermore, since the results come from a man-made non-transparent software package, the correlations between them and physical law are not direct and clear. On the contrary, the analytical methods, concerning the most essential and important elements affecting the physical phenomena *via* ideal assumptions, provide us the most basic law underlying. In addition, the concise and explicit form of exact solution is unmatchable by Monte Carlo simulations. Hence, despite facing tremendous difficulties, the finding of analytical solution of relativistic hydro equations is always our pursuit of the goal. It is an important field in high energy physics.

---
[*] E-mail: jzj265@163.com

The first exact solution of 1+1 hydrodynamics was given by I. M. Khalatnikov about 61 years ago [26], which is for an accelerated system being assumed as a massless ideal fluid and initially at rest. The solution was presented in a complicated integral form and later was used by L. D. Landau in his hydro model study and obtained the rapidity distributions of charged particles [27], which are in generally consistent with the observations made at BNL-RHIC [28-30]. This is the first time for the understanding of the nearly ideal nature of fluid produced in collisions.

The second exact solution of 1+1 dimensional hydrodynamics is given by R. C. Hwa about 41 years ago [31]. This solution is for an accelerationless system with Lorentz invariant initial condition. The result got in this way is simple and explicit. From this solution, J. D. Bjorken was able to get a simple estimate for the initial energy density achieved in collisions from the final observables [32]. This makes the energy density be measurable in experiment. It is the first and by now the only formula being widely recognized as one for estimating the energy density of matter created in collisions. Hence, it receives much attention. This is the reason why Hwa's theory is usually named as Hwa-Bjorken hydro model. However, since the free parameter in the formula has not been well fixed, how to determine the mentioned energy density is still an open problem. Moreover, the invariant rapidity distributions obtained from this model are at variance with experimental observations. Theoretically, such distributions are only the limiting cases of $\sqrt{s_{NN}} \to \infty$.

In recent years, along with the operations of BNL-RHIC and later of CERN-LHC, the investigations of relativistic hydrodynamics have entered into a very active period, becoming one of the most popular subjects. It was during this period that the second and higher order harmonic flows and the ridge structures of the matter created in collisions were observed in experiments [2-6]. This shows us the features of collective flow of produced matter with strong interaction and nearly ideal natures. At the same time, the analytical investigations of hydrodynamics have got into a golden stage of rapid developments and achieved a number of good results [7-15].

For example, by generalizing the relation between ordinary rapidity and space-time one, Ref. [7] integrates Landau and Hwa-Bjorken two famous hydro models into one, becoming a unified hydro model and presenting a set of exact solutions. By taking advantage of the traditional scheme of Khalatnikov potential, Ref. [8] solved analytically the hydro equations and gave a pack of simple exact solutions for an ideal fluid with linear equation of state. By taking into account the work done by the fluid elements on each other, Refs. [9, 10] generalized the Hwa-Bjorken model for an accelerationless system to the model for an accelerated one, and obtained a class of exact analytical solutions of relativistic hydrodynamics.

One of most important applications of 1+1 dimensional hydrodynamics is the analysis of the pseudorapidity distributions of the charged particles produced in nucleus or particle collisions. In our previous work [16], by taking into account the effect of leading particles, we have successfully discussed such distributions for Pb-Pb and Au-Au collisions at respectively CERN-LHC and BNL-RHIC energies in the context of unified hydrodynamics. In this paper, this combined model will be used to analyze the pseudorapidity distributions in the smaller system of Cu-Cu collisions at RHIC energies.

## 2. A brief description of combined model

For the purpose of completeness and application later, we here list the key ingredients of combined model.

(1) The matter created in high energy heavy ion collisions is taken as an ideal fluid fulfilling the equation of state

$$\varepsilon = gp, \tag{1}$$

where $\varepsilon$, $1/\sqrt{g} = c_s$ and $p$ are respectively the energy density, the speed of sound and the pressure of fluid. Investigations have shown that $g$ changes very slowly with energies and centrality cuts [14, 33-35]. For a given incident energy, it can be well taken as a constant.

Eq. (1) allows the expansion of fluid along the colliding axis of two nuclei, that is the longitudinal axis of $z$ to have the form as

$$\begin{aligned}
&\frac{e^{2y}-1}{2}(g+1)\partial_+ p + e^{2y}(g+1)p\partial_+ y + \frac{1-e^{-2y}}{2}(g+1)\partial_- p + \\
&\qquad\qquad e^{-2y}(g+1)p\partial_- y + \partial_+ p - \partial_- p = 0, \\
&\frac{e^{2y}+1}{2}(g+1)\partial_+ p + e^{2y}(g+1)p\partial_+ y + \frac{1+e^{-2y}}{2}(g+1)\partial_- p - \\
&\qquad\qquad e^{-2y}(g+1)p\partial_- y - \partial_+ p - \partial_- p = 0,
\end{aligned} \tag{2}$$

where $y$ is the ordinary rapidity, $\partial_+$ and $\partial_-$ are the compact notation of partial derivatives with respect to light-cone coordinates $z_\pm = t \pm z = x^0 \pm x^1 = \tau e^{\pm \eta_S}$, $\tau = \sqrt{z_+ z_-}$ is the proper time, and $\eta_S = 1/2 \ln(z_+/z_-)$ is the space-time rapidity of fluid.

(2) Eq. (2) is the complicated differential equations with high nonlinearity and coupling between variable $p$ and $y$. In order to solve it, the relation between ordinary rapidity $y$ and space-time $\eta_S$ is generalized to the form [7]

$$2y = \ln u_+ - \ln u_- = \ln F_+(z_+) - \ln F_-(z_-), \tag{3}$$

where $u_\pm = e^{\pm y}$ are the light-cone components of the 4-volicity, and $F_\pm(z_\pm)$ are a priori arbitrary functions. In case of $F_\pm(z_\pm) = z_\pm$, Eq. (3) reduces to $y = \eta_S$, returning to the boost-invariant picture of Hwa-Bjorken. Otherwise, Eq. (3) describes the non-boost-invariant geometry of Landau. Accordingly, Eq. (3) unifies the Landau and Hwa-Bjorken hydrodynamics together. It paves a way between these two models.

By using Eq. (3), Eq. (2) becomes

$$g \partial_+ \ln p = -\frac{(g+1)^2}{4} \frac{f'_+}{f_+} + \frac{g^2-1}{4} \frac{f'_-}{f_+},$$
$$g \partial_- \ln p = -\frac{(g+1)^2}{4} \frac{f'_-}{f_-} + \frac{g^2-1}{4} \frac{f'_+}{f_-}, \tag{4}$$

where $f_\pm = F_\pm/H$, and $H$ is an arbitrary constant. After the above treatments, Eq. (4) becomes solvable. Its solution is [7]

$$s(z_+, z_-) = s_0 \left(\frac{p}{p_0}\right)^{\frac{g}{g+1}} = s_0 \exp\left[-\frac{g+1}{4}(l_+^2 + l_-^2) + \frac{g-1}{2} l_+ l_-\right], \tag{5}$$

where $s$ is the entropy density of fluid, and

$$l_\pm(z_\pm) = \sqrt{\ln f_\pm}, \quad y(z_+, z_-) = \frac{1}{2}(l_+^2 - l_-^2), \quad z_\pm = 2h \int_0^{l_\pm} e^{u^2} du, \tag{6}$$

$h = H/A$, and $A$ is an arbitrary constant.

(3) As the expansion of fluid lasts to the proper time of $\tau_{FO}$, the inelastic interactions between the particles in fluid cease, and the ratios of different kinds of particles maintain unchanged. At this moment, the collective movement of fluid comes to an end, and the fluid decouples or freezes out into the detected particles from a time-like hypersurface with a proper time of $\tau_{FO}$. Considering that the number of charged particles is proportional to the amount of entropy, from solution (5) we can obtain the following rapidity distributions of produced charge particles

$$\frac{dN_{\text{Fluid}}(b,\sqrt{s_{\text{NN}}},y)}{dy} = C(b,\sqrt{s_{\text{NN}}})e^{-(g-1)(l_+-l_-)^2/4} \left.\frac{\partial_+\phi e^y + \partial_-\phi e^{-y}}{\partial_+\phi l_-e^y + \partial_-\phi l_+e^{-y}}\right|_{\tau_{\text{FO}}}, \qquad (7)$$

where $C(b,\sqrt{s_{\text{NN}}})$, independent of rapidity $y$, is an overall normalization constant. $b$ is the impact parameter, and $\sqrt{s_{\text{NN}}}$ is the center-of-mass energy per pair of nucleons. $\phi$ stands for an arbitrary time-like hypersurface.

(4) The right-hand side of Eq. (7) is evaluated on the time-like hypersurface with the proper time equaling $\tau_{\text{FO}}$. Such hypersurface can be therefore taken as

$$\phi(z_+,z_-) = \tau_{\text{FO}}^2 = z_+z_- = C, \qquad (8)$$

where $C$ is an arbitrary constant. This equation gives

$$\partial_\pm\phi = z_\mp. \qquad (9)$$

Thus, Eq. (7) turns into

$$\frac{dN_{\text{Fluid}}(b,\sqrt{s_{\text{NN}}},y)}{dy} = C(b,\sqrt{s_{\text{NN}}})e^{-(g-1)(l_+-l_-)^2/4} \frac{z_-e^y + z_+e^{-y}}{z_-l_-e^y + z_+l_+e^{-y}}. \qquad (10)$$

(5) In nucleus-nucleus collisions, apart from the freeze-out of fluid, leading particles also have certain contribution to the measured charged particles. Leading particles are believed to be formed outside the nucleus, that is, outside the colliding region [36, 37]. The movement and generation of leading particles are therefore free from hydro descriptions. As we have argued before that the rapidity distribution of leading particles takes the Gaussian form

$$\frac{dN_{\text{Lead}}(b,\sqrt{s_{\text{NN}}},y)}{dy} = \frac{N_{\text{Lead}}(b,\sqrt{s_{\text{NN}}})}{\sqrt{2\pi}\sigma}\exp\left\{-\frac{\left[|y|-y_0(b,\sqrt{s_{\text{NN}}})\right]^2}{2\sigma^2}\right\}, \qquad (11)$$

where $N_{\text{Lead}}(b,\sqrt{s_{\text{NN}}})$, $y_0(b,\sqrt{s_{\text{NN}}})$ and $\sigma$ are respectively the number of leading particles, central position and width of distribution. This conclusion comes from the consideration that, for a given incident energy, different leading particles resulting from each time of nucleus-nucleus collisions have approximately the same amount of energy or rapidity. Then, the central limit theorem [38, 39] guarantees the plausibility of above argument. Actually, experimental observations have shown that any kind of charged particles presents a good Gaussian rapidity distribution [28-30]. It is known from Refs. [40-42] that the widths of such Gaussian rapidity distributions increase linearly with beam rapidity. This might indicate the onset of deconfinement from hadron to partonic state even at the CERN-SPS energy scales.

$y_0\left(b,\sqrt{s_{\mathrm{NN}}}\right)$ in Eq. (11) is the average position of leading particles. It should increase with incident energies and centrality cuts. The value of $\sigma$ relies on the relative energy or rapidity differences among leading particles. It should not, at least not apparently depend on the incident energies, centrality cuts and even colliding systems. The concrete values of $y_0$ and $\sigma$ can be determined by tuning the theoretical predictions to experimental data.

By definition, leading particles mean the particles which carry on the quantum numbers of colliding nucleons and take away most part of incident energy. Hence, the number of leading particles is equal to that of participants. For nucleon-nucleon, such as *p-p* collisions, there are only two leading particles. They are separately in projectile and target fragmentation region. For an identical nucleus-nucleus collision, the number of leading particles

$$N_{\mathrm{Lead}}\left(b,\sqrt{s_{\mathrm{NN}}}\right)=\frac{N_{\mathrm{Part}}\left(b,\sqrt{s_{\mathrm{NN}}}\right)}{2}, \quad (12)$$

where $N_{\mathrm{Part}}\left(b,\sqrt{s_{\mathrm{NN}}}\right)$ is the total number of participants, which can be determined in theory by Glauber model [43-45].

## 3. Comparison with experimental measurements

Having at hand the rapidity distributions of Eqs. (10) and (11), the pseudorapidity distributions can be written as [46]

$$\frac{\mathrm{d}N\left(b,\sqrt{s_{\mathrm{NN}}},\eta\right)}{\mathrm{d}\eta}=\sqrt{1-\frac{m^2}{m_{\mathrm{T}}^2\cosh^2 y}}\frac{\mathrm{d}N\left(b,\sqrt{s_{\mathrm{NN}}},y\right)}{\mathrm{d}y}, \quad (13)$$

where $m_{\mathrm{T}}=\sqrt{m^2+p_{\mathrm{T}}^2}$ is the transverse mass, and $p_{\mathrm{T}}$ is the transverse momentum. The first factor on the right-hand side of above equation is actually the Jacobian determinant. This transformation is closed by another relation

$$y=\frac{1}{2}\ln\left[\frac{\sqrt{p_{\mathrm{T}}^2\cosh^2\eta+m^2}+p_{\mathrm{T}}\sinh\eta}{\sqrt{p_{\mathrm{T}}^2\cosh^2\eta+m^2}-p_{\mathrm{T}}\sinh\eta}\right]. \quad (14)$$

Taking into account the contributions from both the freeze-out of fluid and leading particles, the rapidity distributions in Eq. 13 are

$$\frac{\mathrm{d}N\left(b,\sqrt{s_{\mathrm{NN}}},y\right)}{\mathrm{d}y}=\frac{\mathrm{d}N_{\mathrm{Fluid}}\left(b,\sqrt{s_{\mathrm{NN}}},y\right)}{\mathrm{d}y}+\frac{\mathrm{d}N_{\mathrm{Lead}}\left(b,\sqrt{s_{\mathrm{NN}}},y\right)}{\mathrm{d}y}. \quad (15)$$

Inserting above equation or the sum of Eqs. (10) and (11) into (13), we can get the pseudorapidity distributions of charged particles. The results are shown in Figures 1 and 2, which are respectively for distributions in different centrality Cu-Cu collisions at $\sqrt{s_{NN}}$ =200 and 62.4 GeV. The solid dots are the experimental data [47]. The dashed curves are the results obtained from unified hydrodynamics of Eq. (10). The dashed-dotted curves are the results got from leading particles of Eq. (11). The solid curves are the results acquired from Eq. (15), that is, the sums of the dashed and dashed-dotted curves. It can be seen that the theoretical results are in good agreement with experimental measurements.

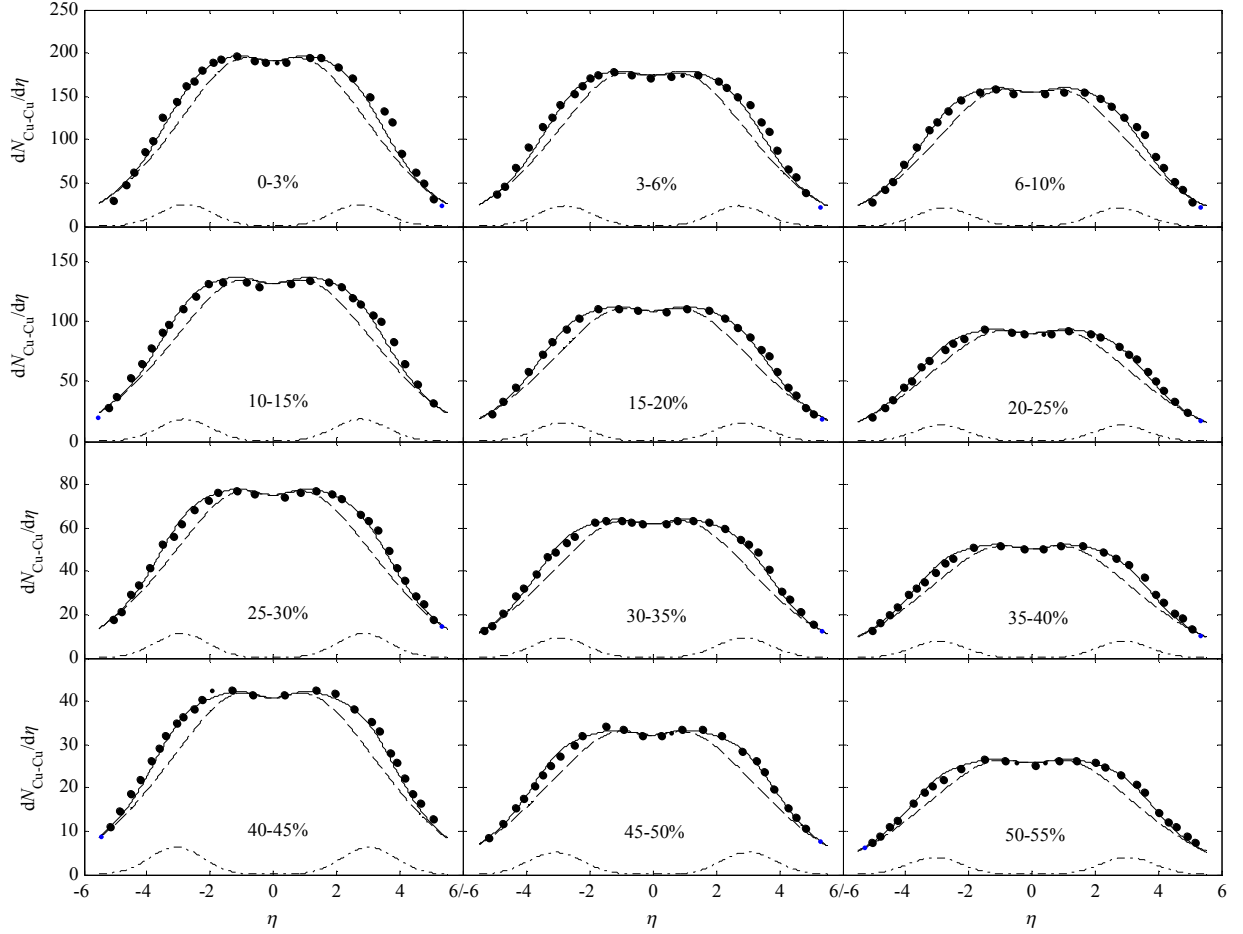

Fig. 1. The pseudorapidity distributions of charged particles produced in different centrality Cu-Cu collisions at $\sqrt{s_{NN}} = 200$ GeV. The solid dots are the experimental measurements [47]. The dashed curves are the results from unified hydrodynamics of Eq. (10). The dashed-dotted curves are the results from leading particles of Eq. (11). The solid curves are the sums of the dashed and dashed-dotted curves.

In calculations, the parameter $g$ in Eq. (10) takes a constant of $g = 8.16$ in both cases of collision energies and in all centrality cuts [33]. The values of $C$ in this equation are summarized in table 1. Here, for the purpose of comparisons, we also list the values of $C$ used in Ref. [16] for Au-Au collisions at $\sqrt{s_{NN}} = 200$ and 62.4 GeV, respectively, as well as the numbers of leading particles evaluated from Eq. (12) for Cu-Cu and Au-Au collisions at the same above stated energies. It can be seen from this table that the variations of $C$ against energies, system sizes and centrality cuts are in just the same way as those of $N_{Lead}$. That is, for a given centrality cut, $C$ increases with energies and system sizes, While, for a given energy and system, $C$ decreases with centrality cuts.

The width parameter $\sigma$ in Eq. (11) takes the value of 0.85 at different incident energies and centrality cuts. As the analyses given above, $\sigma$ is independent of incident energies and centrality cuts. The center parameter $y_0$ in Eq. (11) takes the values of 2.60-2.92 and 2.40-2.49 for centrality cuts from small to large in collisions at $\sqrt{s_{NN}}$ =200 and 62.4 GeV, respectively. As pointed out early, $y_0$ increases with energies and centrality cuts.

Table 1. The values of $C$ and the number of leading particles in different centrality Cu-Cu and Au-Au collisions at $\sqrt{s_{NN}} = 200$ and 62.4 GeV, respectively.

| | Centrality Cuts (%) | | 0-3 | 3-6 | 6-10 | 10-15 | 15-20 | 20-25 | 25-30 | 30-35 | 35-40 | 40-45 | 45-50 | 50-55 |
|---|---|---|---|---|---|---|---|---|---|---|---|---|---|---|
| $C$ | Cu-Cu | 200 GeV | 258.56 | 235.45 | 214.07 | 191.75 | 154.57 | 131.04 | 109.94 | 89.41 | 74.82 | 61.35 | 48.65 | 38.84 |
| | | 62.4 GeV | 134.09 | 126.71 | 120.05 | 104.74 | 90.24 | 76.34 | 63.31 | 50.37 | 41.56 | 34.17 | 28.39 | 22.63 |
| | Au-Au | 200 GeV | 900.43 | 835.44 | 717.95 | 625.55 | 516.32 | 425.14 | 357.09 | 303.18 | 238.11 | 192.00 | 149.04 | ---- |
| | | 62.4 GeV | 485.31 | 465.59 | 423.51 | 362.66 | 300.52 | 250.77 | 197.09 | 167.14 | 128.55 | 107.14 | 83.34 | ---- |
| $N_{Lead}$ | Cu-Cu | 200 GeV | 54.22 | 50.98 | 45.56 | 39.49 | 33.85 | 28.74 | 24.06 | 20.78 | 16.50 | 13.57 | 11.43 | 8.55 |
| | | 62.4 GeV | 53.03 | 48.78 | 44.00 | 38.43 | 32.49 | 26.78 | 22.99 | 19.11 | 16.23 | 13.56 | 10.35 | 8.50 |
| | Au-Au | 200 GeV | 184.41 | 165.52 | 148.67 | 127.49 | 107.67 | 90.05 | 75.12 | 62.86 | 50.74 | 41.61 | 32.47 | ---- |
| | | 62.4 GeV | 178.59 | 159.78 | 142.43 | 120.56 | 101.30 | 82.86 | 70.56 | 56.37 | 46.78 | 39.03 | 28.00 | ---- |

## 4. Conclusions

The matter created in heavy ion collisions is assumed evolving according to the framework of unified hydrodynamics, and then freeze out into charged particles from a pace-like hypersurface with a fixed proper time of $\tau_{FO}$. The typical features of unified hydrodynamics are that: (1) By generalizing the relation between ordinary rapidity $y$ and space-time $\eta_S$, the Hwa-Bjorken and Landau two famous hydro models are integrated as one. (2) In case of linear equation of state, this hydro model can be solved analytically.

In addition to freeze-out of fluid, leading particles also play a certain part to the charged particles. As before, the leading particles are supposed to have Gaussian rapidity distributions normalized to the number of participants, which can be figured out in theory. It is interested to notice that the investigations of present paper once again show that, for a given colliding system, the central position $y_0$ of Gaussian rapidity distribution increases with incident energies and centrality cuts. While, the width parameter $\sigma$ is irrelevant to them, keeping a constant of 0.85. This is consistent with the results arrived at in Ref. [16].

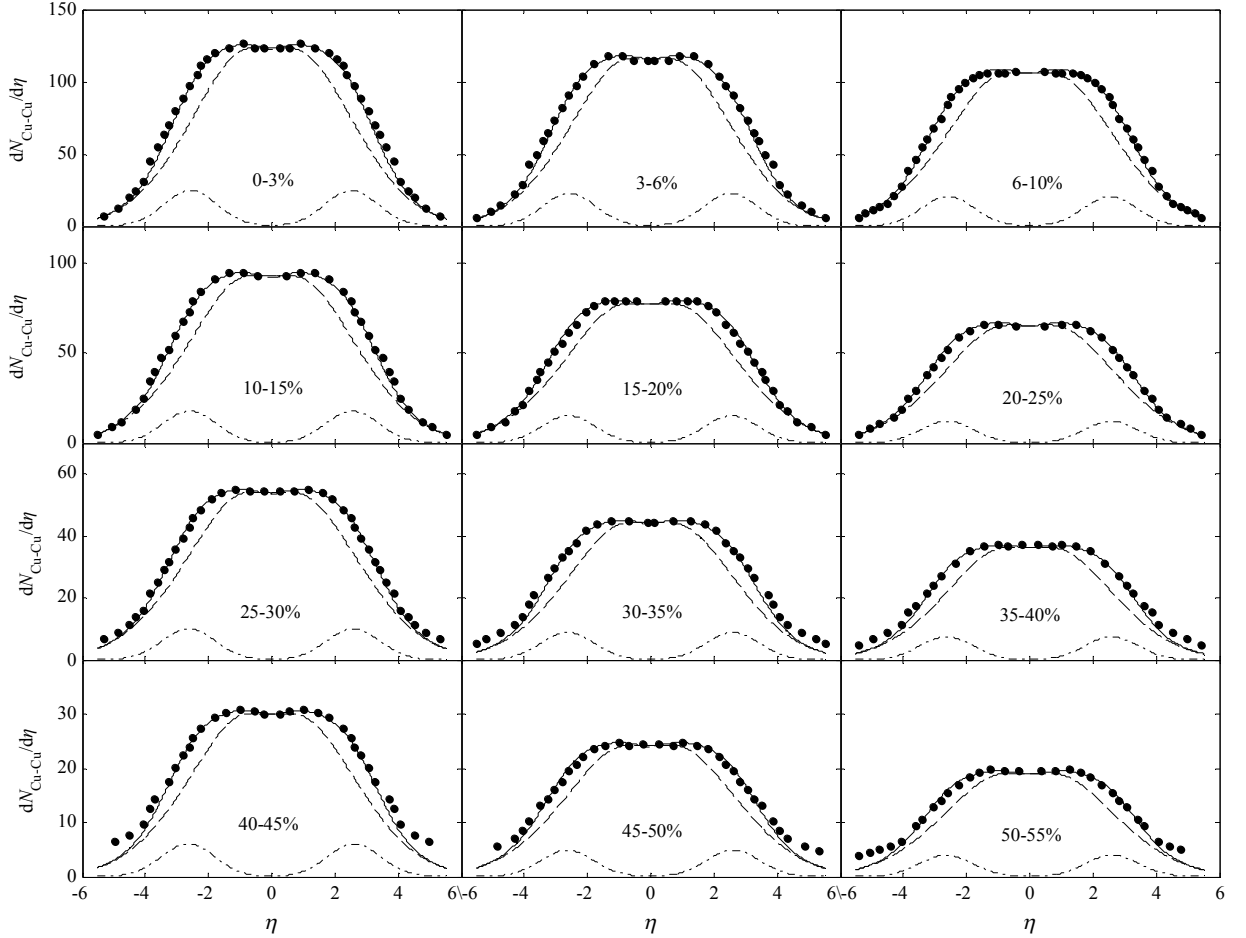

Fig. 2. The pseudorapidity distributions of charged particles produced in different centrality Cu-Cu collisions at $\sqrt{s_{NN}} = 62.4\,\text{GeV}$. The solid dots are the experimental measurements [47]. The dashed curves are the results from unified hydrodynamics of Eq. (10). The dashed-dotted curves are the results from leading particles of Eq. (11). The solid curves are the sums of the dashed and dashed-dotted curves.

Here, it is worth mentioning that the investigations of Refs. [22, 23] also have shown that Landau hydrodynamics alone is not enough in explaining experimental observations in high energy physics. Only after the effects of recombination of constituent quarks in participants are taken into account together, can the experimental measurements in both *p-p* and nucleus-nucleus collisions be described properly. This is in accordance with our analysis. In order to have a good description to experimental data, besides unified hydrodynamics, leading particles are essential as well. Only after the combined effects of them are included simultaneously, can the theoretical predictions match up well with experimental results.

As the end of this paper, we would like to point out that, in recent years, a hydro model named as event-by-event hydrodynamics is widely used in high energy physics [18-21]. This kind of hydrodynamics differs from the one employed in present paper in two main aspects. (1) The former is about the Monte Carlo simulations. The results come from a software package, and the correctness of these results relies on the validity of input parameters. The latter, however, provide us an analytical solution, which is the extraction of the most essential nature of the concerned problem. (2) The former deals with the collisions on a microscopic single event level. Hence, the fluctuating initial conditions are important in explaining the experimental observations, such as, higher flow harmonics and ridge structures. The latter, on the contrary, treats the collisions on the macroscopic "single shot" level with an averaged initial condition. Hence, there is no necessary to consider the fluctuations in initial conditions. This is enough for describing the global variables, such as, the pseudorapidity and transverse momentum distributions.

**Acknowledgments**

This work is partly supported by the Hujiang Foundation of China with Grant No. B14004; the Cultivating Subject of National Project with Grant No. 15HJPY-MS04 and Shanghai Key Lab of Modern Optical System.

**References**

[1] L. D. Landau, *Izvestiya Akademii Nauk SSSR* **17** (1953) 51 (in Russian).

[2] PHENIX Collab. (S. S. Adler *et al.*), *Phys. Rev. Lett*. **91** (2003) 182301.

[3] ALICE Collab. (K. Aamodt *et al.*), *Phys. Rev. Lett*. **107** (2011) 032301.

[4] CMS Collab. (S. Chatrchyan *et al.*), *Phys. Rev. C* **87** (2013) 014902.

[5] CMS Collab. (V. Khachatryan *et al.*), *J. High Energy Phys.* **09** (2010) 091.

[6] STAR Collab. (M. Aggarwal *et al.*), *Phys. Rev. C* **83** (2011) 064905.

[7] A. Bialas, R. A. Janik and R. Peschanski, *Phys. Rev. C* **76** (2007) 054901.

[8] G. Beuf, R. Peschanski and E. N. Saridakis, *Phys. Rev. C* **78** (2008) 064909.

[9] T. Csörgő, M. I. Nagy and M. Csanád, *Phys. Lett. B* **663** (2008) 306.

[10] M. I. Nagy, T. Csörgő and M. Csanád, *Phys. Rev. C* **77** (2008) 024908.

[11] C. Y. Wong, A. Sen, J. Gerhard, G. Torrieri and K. Read, *Phys. Rev. C* **90** (2014) 064907.

[12] M. S. Borshch and V. I. Zhdanov, *SIGMA* **3** (2007) 116.

[13] M. Csanád, M. I. Nagy and S. Lökös, *Eur. Phys. J. A* **48** (2012) 173.

[14] T. Mizoguchi, H. Miyazawa and M. Biyajima, *Eur. Phys. J. A* **40** (2009) 99.

[15] N. Suzuki, *Phys. Rev. C* **81** (2010) 044911.

[16] Z. J. Jiang, J. Wang, K. Ma and H. L. Zhang, *Adv. High Energy Phys.* **2015** (2015) 430606.

[17] Z. J. Jiang, Y. Zhang, H. L. Zhang and H. P. Deng, *Nucl. Phys. A* **941** (2015) 188.

[18] A. Sen, J. Gerhard, G. Torrieri, K. Read and C. Y. Wong, *Phys. Rev. C* **91** (2015) 024901.

[19] G. Gale, S. Jeon and B. Schenke, *Int. J. Mod. Phys. A* **28** (2013) 1340011.

[20] U. Heinz and R. A. Snellings, *Rev. Nucl. Part. Sci.* **63** (2013) 123.

[21] S. Jeon and U. Heinz, *Int. J. Mod. Phys. E* **24** (2015) 1530010.

[22] E. K. G. Sarkisyan and A. S. Sakharov, *Eur. Phys. J. C* **70** (2010) 533.

[23] A. N. Mishra, R. Sahoo, E. K. G. Sarkisyan and A. S. Sakharov, *Eur. Phys. J. C* **74** (2014) 3147.

[24] T. Kalaydzhyan and E. Shuryak, *Phys. Rev. C* **91** (2015) 054913.

[25] Y. Hirono and E. Shuryak, *Phys. Rev. C* **91** (2015) 054915.

[26] I. M. Khalatnikov, *J. Exp. Theor. Phys.* **27** (1954) 529 (in Russian).

[27] S. Z. Belenkij and L. D. Landau, *Usp. Fiz. Nauk* **56** (1955) 309 (in Russian).

[28] BRAHMS Collab. (M. Murray), *J. Phys. G: Nucl. Part. Phys.* **30** (2004) 667.

[29] BRAHMS Collab. (M. Murray), *J. Phys. G: Nucl. Part. Phys.* **35** (2008) 044015.

[30] BRAHMS Collab. (I. G. Bearden *et al.*), *Phys. Rev. Lett.* **94** (2005) 162301.

[31] R. C. Hwa, *Phys. Rev. D* **10** (1974) 2260.

[32] J. D. Bjorken, *Phys. Rev. D* **27** (1983) 140.

[33] PHENIX Collab. (A. Adare *et al.*), *Phys. Rev. Lett.* **98** (2007) 162301.

[34] L. N. Gao, Y. H. Chen, H. R. Wei and F. H. Liu, *Adv. High Energy Phys.* **2014** (2014) 450247.

[35] S. Borsányi, G. Endrődi, Z. Fodor, A. Jakovác, S. D. Katz, S. Krieg, C. Ratti and K. K. Szabó, *J. High Energy Phys.* **77** (2010) 1.

[36] A. Berera, M. Strikman, W. S. Toothacker, W. D. Walker and J. J. Whitmore, *Phys. Lett. B* **403** (1997) 1.

[37] J. J. Ryan, *Proceeding of annual meeting of the division of particles and fields of the APS* (World Scientific, Singapore, 1993), p. 929.

[38] T. B. Li, *The Mathematical Processing of Experiments* (Science Press, Beijing, China, 1980), p. 42 (in Chinese).

[39] J. Voit, *The Statistical Mechanics of Financial Markets* (Springer, Berlin, Germany, 2005), p. 123.

[40] http://arxiv.org/abs/nucl-th/ 611001.

[41] http://arxiv.org/abs/hep-ph/0504207.

[42] http://arxiv.org/abs/hep-ph/0509314.

[43] Z. J. Jiang, *Acta Phys. Sinica*, **56** (2007) 5191 (in Chinese).

[44] Z. J. Jiang, Y. F. Sun and Q. G. Li, *Inter. J. Mod. Phys. E* **21** (2012) 1250002.

[45] Z. W. Wang, Z. J. Jiang and Y. S. Zhang, *J. Univ. Shanghai Sci. Tech.* **31** (2009) 322 (in Chinese).

[46] C. Y. Wong, *Introduction to High Energy Heavy Ion Collisions* (Press of Harbin Technology University, Harbin, China, 2002), p. 23 (in Chinese).

[47] PHOBOS Collab. (B. Alver *et al*.), *Phys. Rev. C* **83** (2011) 024913.